\documentclass{article}
\usepackage{newpasp}
\usepackage{epsfig}
\begin{document}
\title{The VSOP 2 mission: Imaging capabilities}

\author{R. Dodson,
        H. Hirabayashi, Y. Murata, 
        P. G. Edwards,
        K. Wiik}  \affil{Institute of Space and Astronautical Science, 
                 JAXA, 3-1-1 Yoshinodai,
                 Sagamihara, Kanagawa 229-8510, Japan}

\author{D. Murphy}\affil{Jet Propulsion Laboratory, Pasadena, USA}
\author{S. Kameno}\affil{NAO, Mitaka, Tokyo, Japan}

\begin{abstract}

Given the scientific goals of VSOP-2, including the possibility of
observations of the shadows of black holes, we have investigated the
fidelity of the recovered images given a typical {\em uv}-coverage. We
find that we can achieve a dynamic range of better than 1000:1.

\end{abstract}

\section*{Example Science Goals for VSOP-2}

The proposed VSOP-2 mission$^2$ will provide very high resolution at
wavelengths of centimeters and below. VSOP-2, using spacebased mm-VLBI
with full polarisation, will;

\begin{itemize}
\item Image jet formation and acceleration from AGNs.
\item Trace the magnetic field lines in these jets$^6$. 
\item Image accretion disks in nearby AGN black holes$^3$.
\item Image the magnetosphere around proto-stars$^4$.
\item Possibly image the shadow of black holes at a few Schwarzschild Radii
(R$_S$)$^{1,5}$.
\end{itemize}

These science goals, along with the feasible launch parameters, sets
the required orbit for VSOP-2. This sets the required apogee height of
the VSOP-2 orbit to be 25,000 km. This gives, at 43~GHz, a beam size
of 38 $\mu$as (with uniform weighting). This resolution will fulfill
the core science goals.

As shown in Figure 1, the VSOP-2 beam at 43~GHz is smaller than than
10 R$_S$ for Black Holes such as that in M87, allowing the imaging of
the shadow$^5$ and will also allow us to discriminate between the disk
and the core. In Takahashi {\em etal} (2004) general relativistic ray
tracing simulations of the shadow of a black hole are presented, and
he discusses the possibilities of imaging these in detail.

\section*{Imaging Capabilities of VSOP-2}

Given the likely orbital parameters based on a M-V launch (apogee
height of 25,000~km, perigee height of 1,000~km, inclination 31$^o$,
orbital period 7.5~hours) we can calculate the likely {\em
uv}-coverage for all sources in the sky using all possible ground
stations at, for example, 22-GHz. The results for the planned orbital
parameters are shown in Figure 2. The changes in coverage across the
sky can be seen, with gaps in certain alignments and the effect of few
Southern ground stations.

We have made realistic models of the {\em uv}-coverage for the
VSOP-2. We have modelled the the imaging for a likely observing setup,
i.e. the VLBA and some EVN antennae. Our model is a point source and a
Gaussian disk at $40^{\circ}$ (ie., the same declination as the major
VSOP-2 target 3C345). We produced a dirty image from the data and
cleaned and restored in the conventional manner. The dirty and the
restored image are shown in Figure 3.  The restored image has a
dynamic range of greater than 1000. This is with a ``noise free'' data
set, so in practice these levels will be hard to achieve except for
strong sources. Once estimates for likely system temperatures have
been finalised we will produced simulated images which take these
parameters into account.

\section*{Conclusions}

 We have shown that greater than 1000:1 dynamic range VSOP-2 images of
 both resolved and unresolved sources can be obtained in principle
 with the MV-achievable VSOP-2 orbit and using existing ground radio
 telescopes. These are typical structures for the key targets of the
 science goals of the mission. We are developing the skills in
 simulating the capabilities and sensitivities of VLBI using the
 VSOP-2 satellite.

\section*{References}
1 Falcke H., Melia F., Agol E. 2000, ApJ 528, L13\\
2 Hirabayashi, H., 2003, New Technologies in VLBI ed Minh, CS-306\\
3 Manmoto, Mineshige, Kusunose, 1997, ApJ, 554, 964\\
4 Phillips, R. et al 1996, AJ, 111, 918\\
5 Takahashi et al 2004, in prep.\\
6 Zavala, R Taylor, G., 2003, NewAR, 47, 589\\
\\
http://www.vsop.isas.jaxa.jp/vsop2\\
http://vsop.mtk.nao.ac.jp/vsop2\\


\begin{figure}
\begin{center}
\epsfig{file=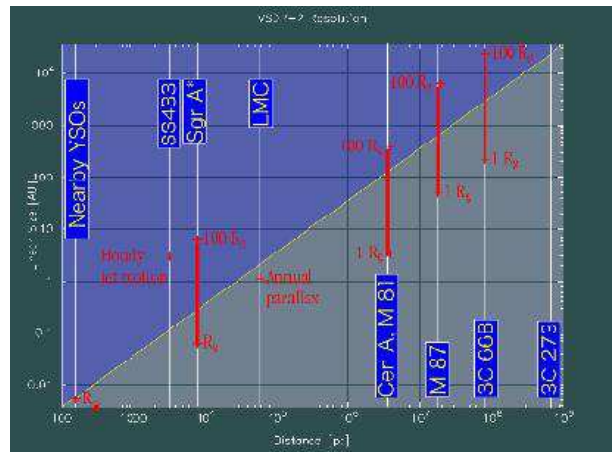,width=8cm}
\caption{VSOP-2 resolution and the range of sizes of interesting
science targets.  The stellar size is shown for nearby Young Stellar
Objects, as is the typical jet motion of the micro quasar SS433. The
event horizon and typical sizes for a disk are shown for a number of
well known black holes.}
\end{center}
\end{figure}

\begin{figure}
\begin{center}
\epsfig{file=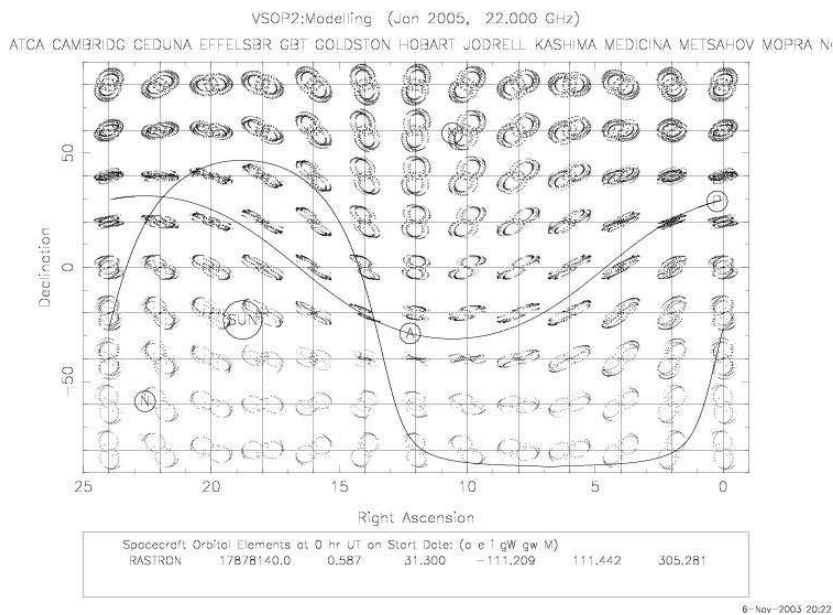,width=8cm,angle=270}
\caption{All sky {\em uv}-coverage for VSOP at 22~GHz. All telescopes
are included in the 12 hour track of sources at the grid
points. Effect of the change in orbital orientation and the small
number of high frequency GRT's in the South can be seen.}
\end{center}
\end{figure}

\begin{figure}
\begin{center}
\epsfig{file=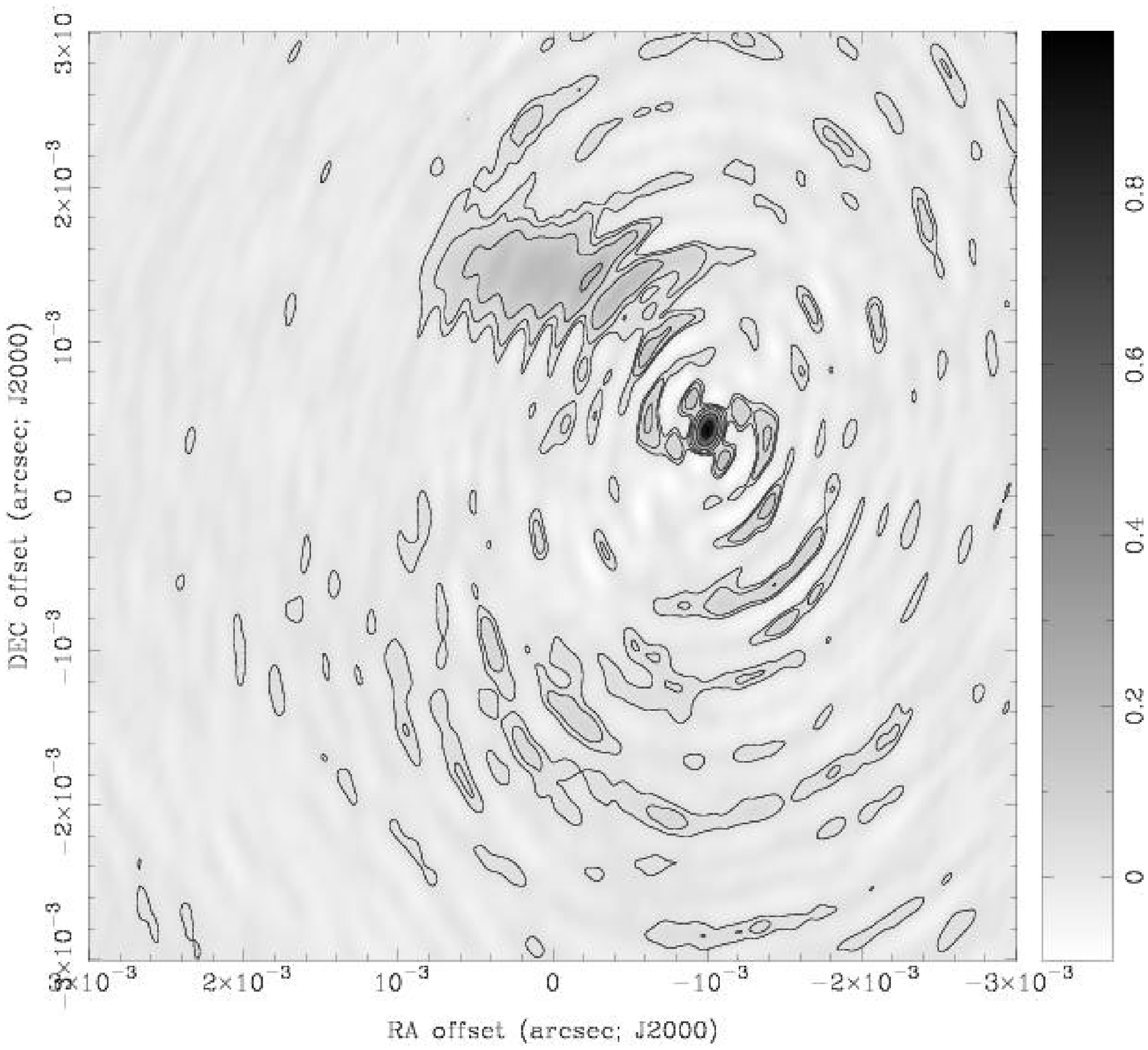,angle=270,width=6cm}
\epsfig{file=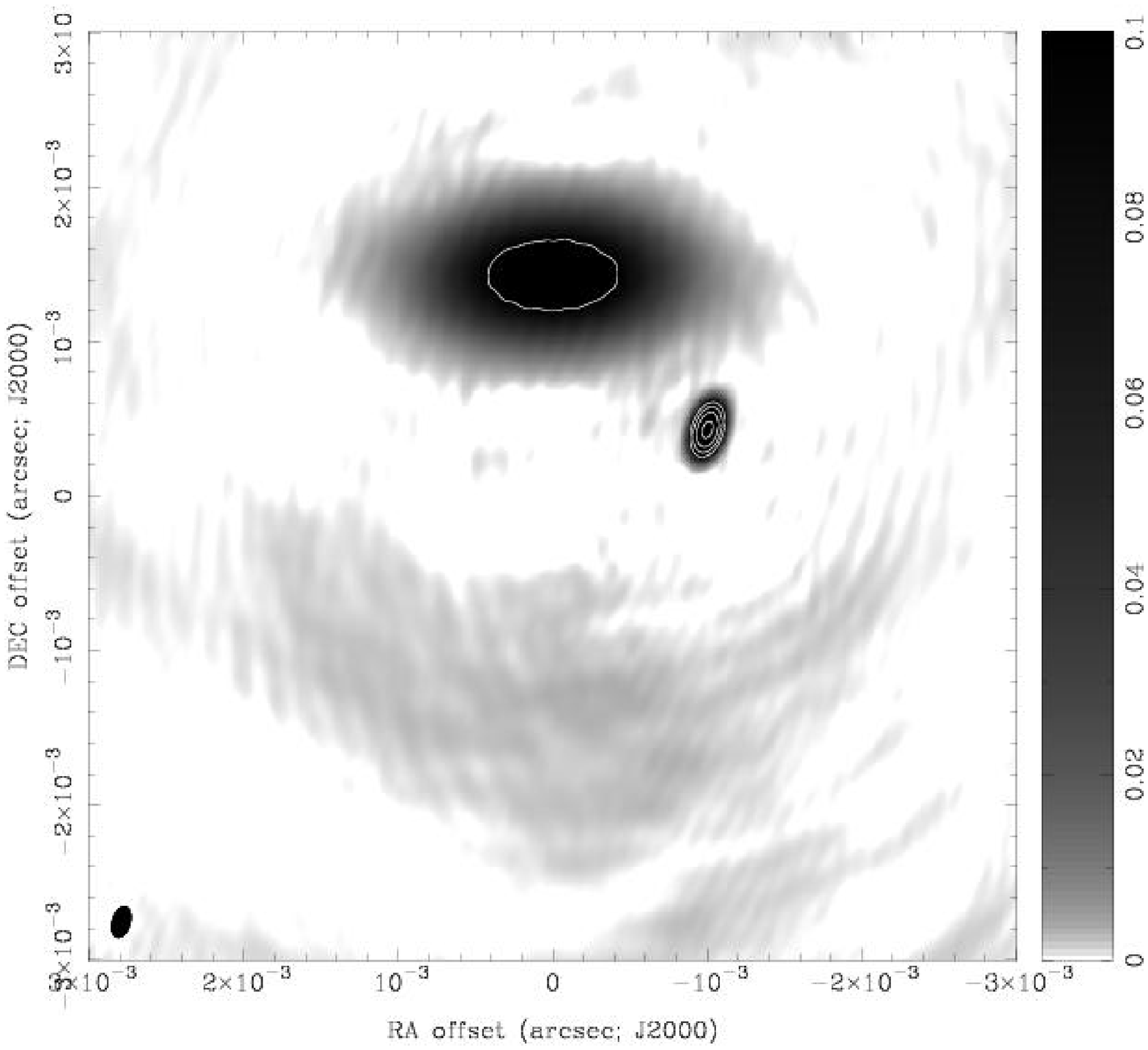,angle=270,width=6cm}
\caption{a) The Dirty image and the b) Restored image of a point
source and a 0.1 by 0.5 mas Gaussian.  The scale is logarithmic over
10\% of the range to highlight the residuals and the contours are at
10, 20, 40 and 80 \% of the peak flux density. The dirty image uses a
linear grey scale and also has the 2.5 and 5\% contours overlaid.}
\end{center}
\end{figure}








\end{document}